# Skyrmion Hall Effect Revealed by Direct Time-Resolved X-Ray Microscopy

Kai Litzius[1,2,3], Ivan Lemesh[4], Benjamin Krüger[1], Pedram Bassirian[1], Lucas Caretta[4], Kornel Richter[1], Felix Büttner[4], Koji Sato[5], Oleg A. Tretiakov[6,7], Johannes Förster[3], Robert M. Reeve[1], Markus Weigand[3], Iuliia Bykova[3], Hermann Stoll[3], Gisela Schütz[3], Geoffrey S. D. Beach[4*], Mathias Kläui[1,2*]

[1]Institute of Physics, Johannes Gutenberg-University Mainz, 55099 Mainz, Germany

[2]Graduate School of Excellence Materials Science in Mainz, 55128 Mainz, Germany

[3]Max Planck Institute for Intelligent Systems, 70569 Stuttgart, Germany

[4]Department of Materials Science and Engineering, Massachusetts Institute of Technology, Cambridge, Massachusetts 02139, USA

[5]WPI Advanced Institute for Materials Research, Tohoku University, Sendai 980-8577, Japan

[6]Institute for Materials Research, Tohoku University, Sendai 980-8577, Japan

[7]School of Natural Sciences, Far Eastern Federal University, Vladivostok 690950, Russia

* Authors to whom correspondence should be addressed: klaeui@uni-mainz.de, gbeach@mit.edu

**Magnetic skyrmions are highly promising candidates for future spintronic applications such as skyrmion racetrack memories and logic devices. They exhibit exotic and complex dynamics governed by topology and are less influenced by defects, such as edge roughness, than conventionally used domain walls. In particular, their finite topological charge leads to a predicted "skyrmion Hall effect", in which current-driven skyrmions acquire a transverse velocity component analogous to charged particles in the conventional Hall effect. Here, we present nanoscale pump-probe imaging that for the first time reveals the real-time dynamics of skyrmions driven by current-induced spin orbit torque (SOT). We find that skyrmions move at a well–defined angle $\Theta_{SH}$ that can exceed 30º with respect to the current flow, but in contrast to theoretical expectations, $\Theta_{SH}$ increases linearly with velocity up to at least 100 ms$^{-1}$. We explain our observation based on internal mode excitations in combination with a field-like SOT, showing that one must go beyond the usual rigid skyrmion description to unravel the dynamics.**

In most magnetic materials, the exchange interaction is symmetric, favoring a collinear alignment of spins. However, recently a new type of magnetic system with broken inversion symmetry has moved to the focus of modern solid-state physics. In such systems, the interfacial Dzyaloshinskii-Moriya interaction (DMI)[1,2] favors a perpendicular alignment of adjacent spins and its interplay with the symmetric exchange interaction then leads to a helical spin structure with well-defined handedness.[3-9] One of the most interesting spin structures that originates from this type of material is the magnetic skyrmion.[3,4,10,11] This topologically non-trivial, particle-like structure exhibits very interesting properties, such as topological stabilization and predicted efficient current-induced motion.[12-15] One major advantage is that skyrmions experience a repulsive force from geometrical boundaries such as the edges of a magnetic track. This is in contrast to ordinary domain walls, which are strongly affected by edge roughness. As a result, skyrmions interact less with geometrical irregularities and are expected to move smoothly and with high stability along the track. In combination with an efficient driving mechanism, such as spin orbit torques,[16-18] this makes them promising candidates for future applications, where fast and reliable switching is key.[19,20] However, during current-induced displacements, skyrmions have been predicted to exhibit an effect that is similar to the Hall effect for electrons and that is therefore called skyrmion Hall effect. As a result, the current not only induces longitudinal but also transversal motion of the skyrmions, which has to be considered especially in logic gates.[20] The skyrmion Hall effect plays an important role in the dynamics of skyrmions because it can change the transversal position of the skyrmion within the track and therefore lead to the annihilation of the skyrmion quasi-particle at the edge. Since skyrmions at rest are expected to be always centered in the track due to the edge repulsion, quasi-static imaging can lead, in particular in low-pinning systems, to incorrect determination of the skyrmion Hall angle. Only direct dynamic imaging of the displacement in real time can reveal the dynamic velocities and displacement direction to understand and tailor the skyrmion motion in devices.

In this work, we perform pump-probe x-ray microscopy that for the first time reveals the real-space trajectories and dynamic velocities of individual skyrmions in a magnetic racetrack. Our results demonstrate that the necessary reproducibility and uniformity of the motion for time resolved measurements of current induced dynamics excited by spin orbit torques (SOTs) is indeed realized in our optimized stack structure, showing the applicability for devices. We observe a large skyrmion Hall angle and find a strong dependence on the skyrmion velocity that cannot be explained by standard micromagnetic simulations that only include the conventionally used damping-like (DL)-SOT.[16] We find that micromagnetic predictions qualitatively reproduce the observed skyrmion behavior when including field-like (FL) SOTs. This result suggests that the FL-SOT plays a more important role than expected so far and we explain the observations based on excitations of the skyrmion spin structure that go beyond the previously assumed rigid skyrmion description of the dynamics.

To reveal the interplay between FL and DL torques and their influence on the resulting skyrmion dynamics, we used an optimized low pinning stack based on thin multilayers of [Pt(4.5 nm)/CoFeB(0.7 nm)/MgO(1.4 nm)]$_{15}$. The contrast between very high DMI at the Pt/CoFeB interface and the lower DMI at the CoFeB/MgO interface results in a large net DMI and additionally, the materials composition yields a strong perpendicular magnetic anisotropy (PMA). Together, these are able to stabilize magnetic skyrmions in the stack at room temperature as previously shown in Ref. [17]. Furthermore, this material exhibits excellent homogeneity, resulting in a very low pinning energy landscape and thus has great potential for obtaining reliable skyrmion dynamics. In particular, the reproducibility of skyrmion motion resulting from low pinning is crucial for pump-probe dynamics measurements that reveal the dynamic velocities. Since the experimental contrast is integrated over billions of repetitions of the excitation process, variations in the skyrmion trajectories due to stochastic processes would not allow for a clear signal. The schematics of the measurement can be found in Figure 1.

First, we determine the topological properties and average velocities of the skyrmions, which identify the observed spin structures as chiral skyrmions and not non-chiral bubble skyrmions.[13,21] As already introduced in Ref. [17], we used the dependence of the skyrmion trajectory on the DMI to determine the chirality of the system. The direction of the SOT driven motion of the skyrmion along the track (± x) depends only on the sign of the DMI constant D, for a given spin Hall angle, and is independent of the skyrmion polarity. We find that the skyrmion motion occurs against the electron flow direction, which is the characteristic behavior for left-handed Neél domain walls and skyrmions, as expected in Pt interface based materials when Pt is the bottom interface. Having established the topological properties, we turn to the dynamics in more detail. First we apply an out-of-plane (OOP) magnetic field of 30 mT to set the size of the skyrmions [11,17]. A typical view of such a skyrmion and a fit of its out-of-plane (OOP) magnetization profile can be found in Figure 2A+B. Note that the analysis takes into account the convolution with the x-ray beam diameter that increases the observed contrast diameter as compared to the real skyrmion diameter. The skyrmion diameter shown in the figure represents the apparent skyrmion size during the dynamics. To excite the dynamics, we applied bipolar current pulses designed to drive the skyrmions back and forth to restore the initial state after each pump-probe cycle. This allows for time-resolved imaging where each frame corresponds to a snapshot of the skyrmions during the measurement cycle and therefore also during the excitation (details in methods section).

Previous work demonstrated high average skyrmion velocities exceeding 100 ms$^{-1}$ at current densities around 5×10$^{11}$ Am$^{-2}$, enabled by the low pinning in this material[17]. However, that work relied on static imaging of the skyrmion positions before and after current-pulse application, without providing information on the dynamic trajectories and velocities. Pump-probe measurements, by contrast, provide access to dynamic (time-resolved) velocity profiles of the skyrmions, which cannot be directly inferred from conventional static measurements because material inhomogeneities (e.g. pinning sites) can significantly influence the dynamic trajectory and velocities. Furthermore, the open question of the effective mass and the inertia, as previously found to be sizeable for bubble skyrmions[13], can only be revealed by dynamic imaging. Therefore, as a first step, we extract the dynamic velocities of the skyrmions as a function of time. This can be done by tracking the skyrmion positions in each frame of the pump-probe measurement and then calculating the displacements within the given time between the frames. Because our samples are placed on thin silicon nitride membranes in vacuum, the heat dissipation is poor. For this reason, we include an increased delay between the unipolar pulses with positive and negative polarity to allow the system to cool down (in the supplemental movies we omit the frames corresponding to this delay period to highlight the important times during the pulse injection where the dynamics occur). Figure 2C shows the time resolved skyrmion velocity. It can be seen that the velocity follows the current density without any noticeable delay, implying that the observed skyrmions exhibit only a small inertia (upper bound ≈ 1.3×10$^{-21}$ kg), which is significantly lower than the values measured in [13], where the inertia was found to be large for non-chiral bubble skyrmions. This can be explained as derived in [13]: The skyrmion mass scales inversely with the rigidity of the spin structure, which in our case is very high because of the strong DMI. The DMI lowers the effective mass, because it makes the spin structure more rigid and thus counteracts the deformations needed for generating the effective mass. As a result, the mass of our skyrmions can be expected to be small as compared to the non-chiral bubble skyrmions observed in [13], which is in line with the observations and in agreement with our micromagnetic simulations showing a delay of < 100 ps and a very similar response as compared to the experimental observations (Fig. S2).

Furthermore, the extracted peak dynamic velocities are slightly higher than the averaged ones in the current density versus velocity plots derived from static measurements[17]. This effect stems from the unknown threshold current density in the static imaging and the dynamic velocity vs. current-density dependence that only dynamic imaging is able to reveal. Comparing the width of the pulse and the width of the response, one can see that the absolute amount of time the skyrmions move due to the current pulse (~6 ns) is comparable to the full-width-at-half-maximum (FWHM) value of the excitation itself, while the shapes fit the expected ones from simulations very well (Fig. S2). This means the skyrmions move very efficiently, with velocities reaching those predicted for perfect, pinning-free, systems. The knowledge of the exact velocity as a function of current density is especially important for short current pulses with non-rectangular shape. In this case, the skyrmions do not move with the velocities expected from averaged current density values and therefore will end up at a different location in the track, which can only be predicted by knowing the dynamic velocities that we reveal here.

Next, we use dynamic imaging to study the skyrmion properties during the dynamics, which is completely inaccessible by conventional quasi-static measurements. Micromagnetic simulations predict that the skyrmion can change its size and in particular above a threshold current density, the skyrmion size is predicted increases significantly and continuously during its motion. This can significantly alter the skyrmion's behavior and, since it shrinks back to its equilibrium size after several nanoseconds, cannot be detected without dynamic imaging. For our current densities our experiment reveals no significant difference of the size due to the current excitation within the spatial resolution of 20 nm, showing that the skyrmions exhibit stable dynamics with no significant increase in diameter.

Finally, we analyze the skyrmion trajectory and compare the direction of the motion to the applied current flow direction. To do this we plot parallel and perpendicular components of the trajectories of several skyrmions visible within the STXM field-of-view, where the current-flow is directed along the x-axis. The trajectories are shown for one of the different experimentally investigated current densities (4.2×10$^{11}$ Am$^{-2}$) (Fig. 2D). The parallel trajectories of several freely moving skyrmions are clearly visible, indicating fully reproducible motion without significant influence from pinning (otherwise the pump-probe measurements which combine over 10$^9$ repetitions of the dynamic process would only show noise due to stochastic motion). All skyrmion trajectories show the same constant angle with respect to the current flow direction for the same current density, which is hallmark of the skyrmion Hall effect.

Surprisingly, our experimental data collected for several current densities shows a pronounced dependence of the skyrmion Hall angle on the skyrmion velocity as shown in Figure 3. This is in contrast to previous micromagnetic simulations,[22] as well as analytical models of rigid skyrmion motion driven by antidamping torque,[16] which predict a velocity-independent skyrmion Hall angle Θ$_{SH}$ given by $\tan \Theta_{SH} = G/\alpha \tilde{D}$, where $G$ is the gyrovector, $\alpha$ is the damping parameter, and $\tilde{D}$ is the dissipation factor[16]. For a skyrmion whose radius $R$ is much larger than the domain wall width $\Delta$, one can show that $\tan \Theta_{SH} \approx \pm 2\Delta/\alpha R$ using the analytical expression for the skyrmion spin texture in Ref. 23. For $\Delta \approx 15 \text{ nm}$ estimated from our material parameters and in line with micromagnetic simulations, conventional models predict a velocity-independent $\Theta_{SH} \approx 50°$ assuming α=0.5, which is in good agreement with our micromagnetic results based on damping-like (DL) spin-orbit torque (SOT).

Experimentally, the absolute values of the skyrmion Hall angles are smaller than the micromagnetic predictions, and we find a linear dependence of Θ$_{SH}$ on the velocity (current density) as seen in Fig. 3. These behaviors cannot be explained by conventional micromagnetic simulations including only a DL-SOT. In addition, for reasonable changes of material parameters such as the DMI constant, saturation magnetization M$_s$ or damping parameter which could give an increased error bar (see Fig. S3) we also cannot explain the behavior (see supplementary information section S3) at the experimental current densities.[‡]

We conclude further that there must be an additional, intrinsic effect, which is influencing the skyrmion trajectories. We have studied possible scenarios that can explain this increase of the observed angles with increasing skyrmion velocity. One parameter, which has previously been neglected, is the sizeable field-like (FL) SOT that has been observed in these multilayer materials. Its influence can be quantified by the field-like parameter $\xi$, which is the ratio of the FL and DL torque. So far, only rigid skyrmions or skyrmions that only deform slightly and excitations with a simplified DL-SOT have been considered. In this case, simulations and the Thiele equation assuming a fully rigid skyrmion predict no influence of the FL-SOT on the skyrmion trajectories at all, whereas additional degrees of freedom that allow the skyrmion to slightly deform and breathe[26] can lead to an influence of a FL-SOT as we show here (Fig. 4, more details in Fig. S4 and Fig. S5).[27] Together with a non-simplified SOT with DL and FL components[28-30] we find a significant influence of the FL-SOTs: an increasing current and thus an increasing FL-SOT changes the skyrmion Hall angle (Fig. 4). It furthermore affects larger or faster skyrmions to a much greater extent than smaller or slower ones since the former are less rigid and thus are more susceptible to this effect. While the calculated and observed skyrmion Hall angle dependence on the current density agree qualitatively, we find some quantitative differences. These can be attributed to thermal excitations, which increase the skyrmion deformation but are not taken into account in the zero temperature simulations, calling for future theoretical efforts beyond the scope of this work.

The displacement of skyrmions occurs at an angle with respect to the current flow direction demonstrating the presence of a skyrmion Hall effect. This angle increases linearly with the velocity and we can qualitatively attribute this to the previously underestimated influence of the field-like SOT for skyrmions that deform during the dynamics. This offers a new approach to explain the complex skyrmion dynamics and provides an additional handle for tailoring the dynamic properties. Combined with the highly reproducible dynamics, we thus show that the developed Pt/CoFeB/MgO multilayer stack exhibits skyrmions that are apt for spintronics applications.

---

[‡] We note that during the preparation of this manuscript, we became aware of a related static observation of the skyrmion Hall angle as a function of current density and a corresponding theoretical work [24,25] showing qualitatively a similar trend at low current densities that could originate from pinning.

**Methods**

We have grown multilayer films of [Pt(4.5 nm)/CoFeB(0.7 nm)/MgO(1.4 nm)]$_{15}$ using dc and rf magnetron sputter deposition at room temperature under an argon pressure of 3mTorr and at a background pressure of ~$2.0 \times 10^{-7}$ Torr. A thin underlayer of Ta (3 nm) was used to improve adhesion between film and substrate. Films were deposited on two types of substrates: on 100 nm and 200 nm-thick Si$_3$N$_4$ membranes used for XMCD imaging as well as on thermally oxidized Si wafers used for characterization with vibrating sample magnetometry (VSM).

VSM measurements of the multilayer stack resulted in a saturation magnetization Ms of $4.3 \times 10^5$ A/m (per CoFeB volume) and an in-plane saturation field $\mu_0 H_k$ of 0.7 T.

The track structures were patterned with electron beam lithography and lift-off process. Contact pads were patterned with a second lithography step, followed by the deposition of Ti(5nm)/Au(100nm) bilayers using e-beam evaporator.

The measurements were carried out using scanning transmission x-ray microscopy (STXM) at the MAXYMUS beamline at the BESSY II synchrotron in Berlin, Germany. This technique exploits the X-ray Magnetic Circular Dichroism (XMCD) effect that leads to a different absorption of circular polarized X-rays dependent on the magnetization structure in the sample. The result is a difference in the transmitted photon intensity, which directly corresponds to the magnetization pattern within the sample (Fig. 1B). The time-resolved measurements were performed in pump-probe mode of the setup, allowing us to synchronize the excitation of the sample (pump) with the illuminating X-ray bunches of the synchrotron (probe). This finally ends up in a stroboscopic movie that shows the time-resolved response of the sample in frames. Imaging was performed in out-of-plane (OOP) mode, resulting in an orthogonal incidence angle of the circularly polarized x-ray beam that probes the OOP component of the magnetization in the film. Since the technique uses transmitted photons to detect the magnetization pattern, the information about magnetization structures in the film is intrinsically averaged over all layer repetitions.

The Pt/CoFeB/MgO devices were 2 μm wide and 5 μm long. Pump-probe measurements were performed by injecting bipolar current pulses into the track through gold contacts deposited on top. The duty cycle was chosen to be as low as possible in order to give the system enough time to cool down during the measurements. The repetition rate of the applied bipolar pulses was ~750 kHz and the length (FWHM) of a single pulse ~5 ns. The mentioned current densities were calculated under the assumption that the amount of current in the MgO layer can be neglected; that is, only the thicknesses of Pt and CoFeB were used to determine the current densities.

The positions of the centers of the skyrmions $x_i$ and $y_i$ were calculated by their center-of-mass with the weights ($c - p_i$), where $p_i$ is the grey value of the corresponding pixel (ranging from 0 to 255) and c is the highest possible grey-value (255). The center of mass of each skyrmion was calculated within the circular mask with a radius of 5 pixels (for the particular field of view and pixel resolution chosen), which ensures that each skyrmion is fully inside the mask. In order to minimize the error from the position of the mask with respect to the skyrmion, we shifted the mask to the four nearest neighboring pixels and the resulting skyrmion position was estimated by averaging the five center positions to achieve maximal accuracy.

Micromagnetic simulations were performed based on the MicroMagnum micromagnetic software, available at http://micromagnum.informatik.uni-hamburg.de/ with additional modules for DMI and SOTs developed in our group. To speed up the simulations, we used the effective medium approach introduced in [17]. This approach can shrink the whole film with 45 layers into one effective layer, which exhibits the same static and dynamic properties as the full film. The systems were discretized with mesh sizes of 1×1 nm² and scaled according to the effective medium model. The simulations used a saturation magnetization of 4.3×10⁵ A/m, an exchange constant $A$ = 1.0×10⁻¹¹ J/m, a Gilbert damping of 0.5 due to the Pt layer, and an out-of-plane uniaxial anisotropy with in-plane saturation field $\mu_0 H_k$ = 0.7 T, corresponding to the values measured by VSM. The SOT was included with spin Hall angle $\alpha_H$ = +0.15 and a |ξ| of up to 5. All values are given as measured, before applying the effective scaling. The centers of the simulated skyrmions were calculated by the center-of-mass with the weights $\frac{M_{z,i} - M_S}{2}$, where $M_{z,i}$ is the z-component of the magnetization in cell i and $M_S$ the saturation magnetization. In agreement with the experimental description given in [30], the SOTs were implemented by adding the torques directly to the explicit LLG equation:

$$\frac{d\vec{M}}{dt} = -\frac{\gamma_0}{1+\alpha^2}\vec{M} \times \vec{H}_{eff} - \frac{\gamma\alpha}{(1+\alpha^2)M_S}\vec{M} \times (\vec{M} \times \vec{H}_{eff}) + \tau_{SOT}$$

The SO torque term $\tau_{SOT}$ then enters as

$$\tau_{SOT} = \frac{\gamma_0}{1+\alpha^2} \cdot a_j \cdot \left((1+\xi\alpha)(\vec{M} \times (\vec{M} \times \vec{p})) + (\xi - \alpha)M_S(\vec{M} \times \vec{p})\right)$$

where $\vec{p} = \vec{e}_J \times \vec{e}_z$ is the spin polarization generated by the technical current J and $a_j = \frac{\hbar}{2\mu_0 e}\frac{\alpha_H |J|}{M_S^2 d}$ is the spin Hall parameter, composed by the spin Hall angle $\alpha_H$, the amplitude of the applied current, the saturation magnetization $M_S$ and the thickness $d$ of the magnetic layer. The field-like pre-factor is then effectively given by $\xi a_j$.

**Acknowledgments:**

Work at MIT was primarily supported by the U.S. Department of Energy (DOE), Office of Science, Basic Energy Sciences (BES) under Award #DE-SC0012371 (sample fabrication). G.B. acknowledges support from C-SPIN, one of the six SRC STARnet Centers, sponsored by MARCO and DARPA. M.K. and the group at Mainz acknowledge support by the DFG (in particular SFB TRR173 Spin+X) the Graduate School of Excellence Materials Science in Mainz (MAINZ, GSC 266) the EU (MultiRev (ERC-2014-PoC 665672), MASPIC (ERC-2007-StG 208162), WALL, FP7-PEOPLE-2013-ITN 608031), MOGON (ZDV Mainz computing center) and the Research Center of Innovative and Emerging Materials at Johannes Gutenberg University (CINEMA). B.K. is grateful for financial support by the Carl-Zeiss-Foundation. O.A.T. acknowledges support by the Grants-in-Aid for Scientific Research (Grants No. 25800184, No. 25247056, and No. 15H01009) from the Ministry of Education, Culture, Sports, Science and Technology (MEXT) of Japan and SpinNet. K.L. gratefully acknowledges financial support by the Graduate School of Excellence Materials Science in Mainz (MAINZ) and the help and advice of Karin Everschor-Sitte and technicians of the Kläui group, especially Stefan Kauschke. Measurements were carried out at the MAXYMUS end station at Helmholtz-Zentrum Berlin. We thank HZB for the allocation of beamtime.

**Author Contributions**

M.K. and G.S.D.B. proposed and supervised the study. I.L., L.C., and K.L. fabricated devices. I.L. and L.C. performed the film characterization. K.L., L.C., K.R., P.B., J.F., R.R., H.S., G.S., J.B., and M.W. conducted STXM experiments at MAXYMUS beamline at the BESSY II synchrotron in Berlin. K.L. and M.K. performed and analyzed the micromagnetic simulations. B.K., K.S., and O.A.T. derived a Thiele equation to explain the micromagnetic simulations and provided input for the latter. K.L, P.B., and K.R. performed the analytical analysis of the experimental data. F.B. derived the expression for the skyrmion Hall angle as function of the domain wall width. All authors participated in the discussion and interpreted results. K.L. drafted the manuscript with the help of M.K. and assistance from G.S.D.B. All authors commented on the manuscript.

**Author Information**

The authors declare no competing financial interests.  Correspondence and requests for materials should be addressed to M.K. (klaeui@uni-mainz.de) and G.S.D.B. (gbeach@mit.edu).

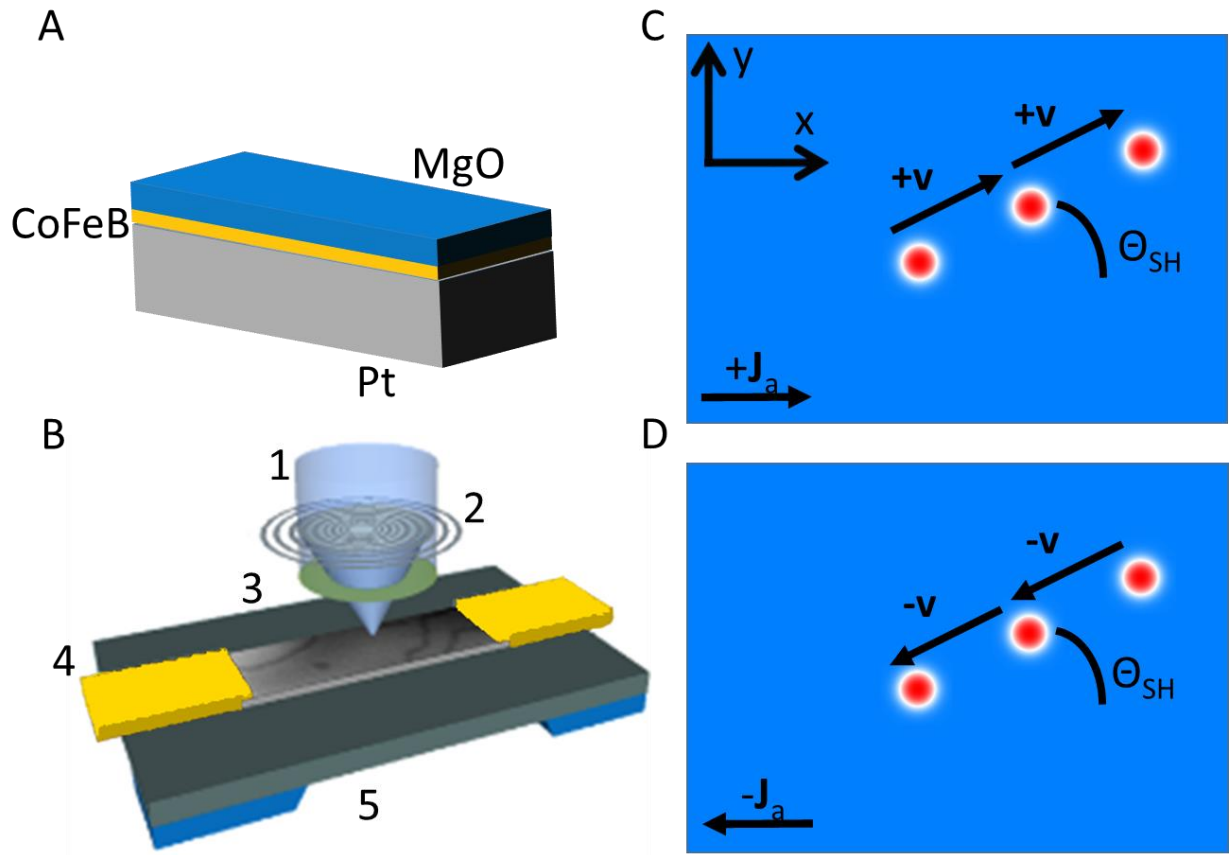

**Figure 1: Schematic description of technique and observed skyrmion Hall effect.** A) Observed stack as single repetition. The layers show the relative experimental thickness differences. B) Schematics of the Scanning Transmission X-ray Microscopy (STXM) measurements. A x-ray beam (1) is focused via a zone plate (2) and an apperture (3) on the sample (4), which is contacted by two gold striplines. The beam transmits through the sample and the substrate's silicon nitride membrane (5). On the other side (not shown), the beam is detected. The transmission of x-rays is dependent on the magnetization at the specific focus spot. C+D) Skyrmions moving in a magnetic wire (micromagnetic simulations, red and blue correspond to positive and negative out-of-plane contrast, respectively). When excited by a alternating current, the skyrmions move back-and forwards with a specific angle with respect to the current flow direction. This is the skyrmion Hall angle $\Theta_{SH}$.

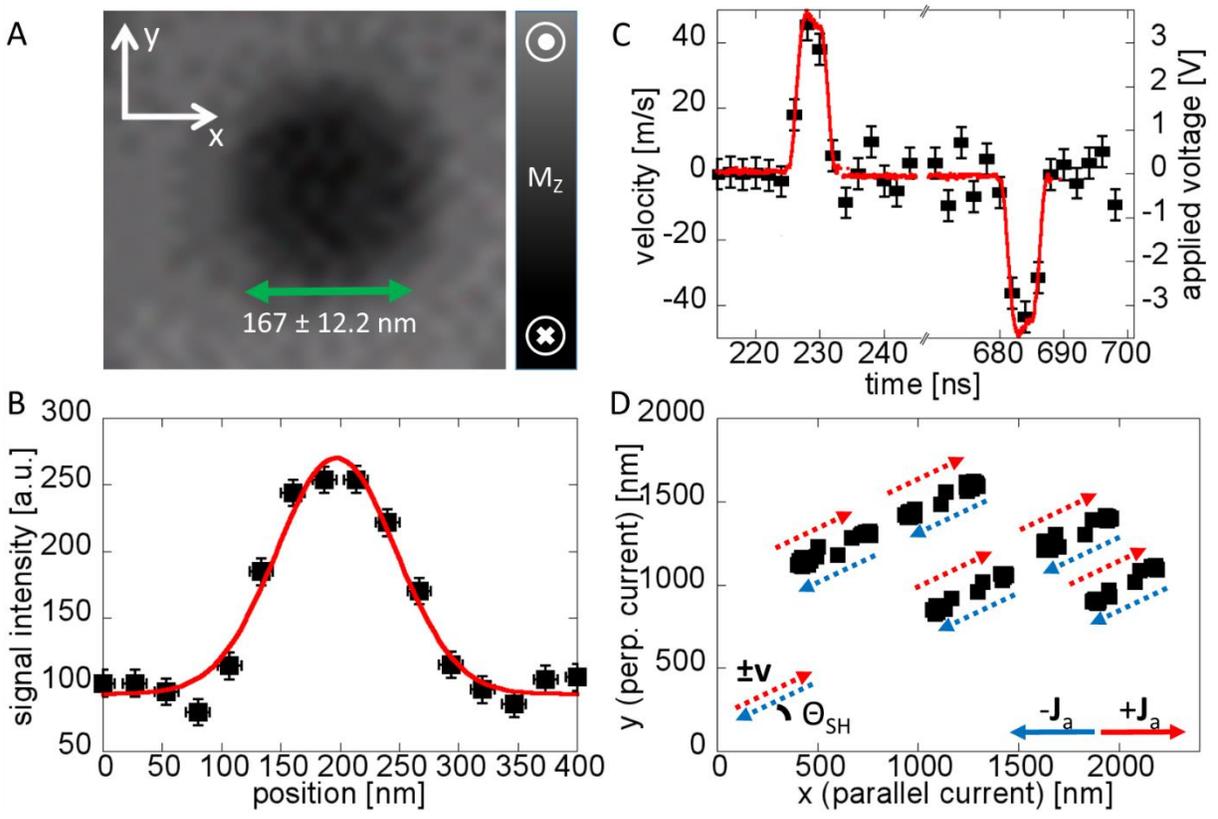

**Figure 2: Analysis of skyrmions and skyrmion trajectories.** A+B) Image of a typical skyrmion as seen by STXM imaging with the size indicated (excited size during the dynamics) and fitted profile. C) Dynamic velocity as a function of time (black) with applied voltage pulse indictated (red). No visible delay implies low inertia and thus a low effective mass. D) Skyrmion trajectories for one experimental current density ($4.2 \times 10^{11}$ Am$^{-2}$). It is apparent that all skyrmions move in parallel and synchronously. The vertical range of 2 µm corresponds to the width of the used wire.

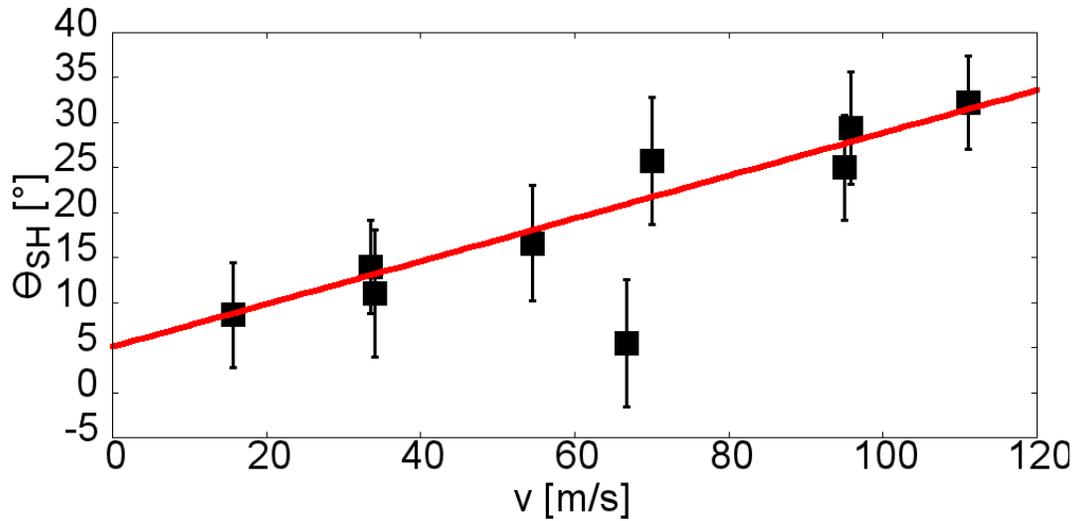

**Figure 3. Experimentally observed "skyrmion Hall angles" of the skyrmion displacement direction with respect to the current flow direction for different velocities.** When plotted as a function of the extracted dynamic peak-velocities of the observed angles a linear dependence is found. This is consistent for all samples, even with different pinning properties (different threshold current densities to move the skyrmions), implying that pinning cannot be the only origin of this behavior. The linear fit is a guide for the eye.

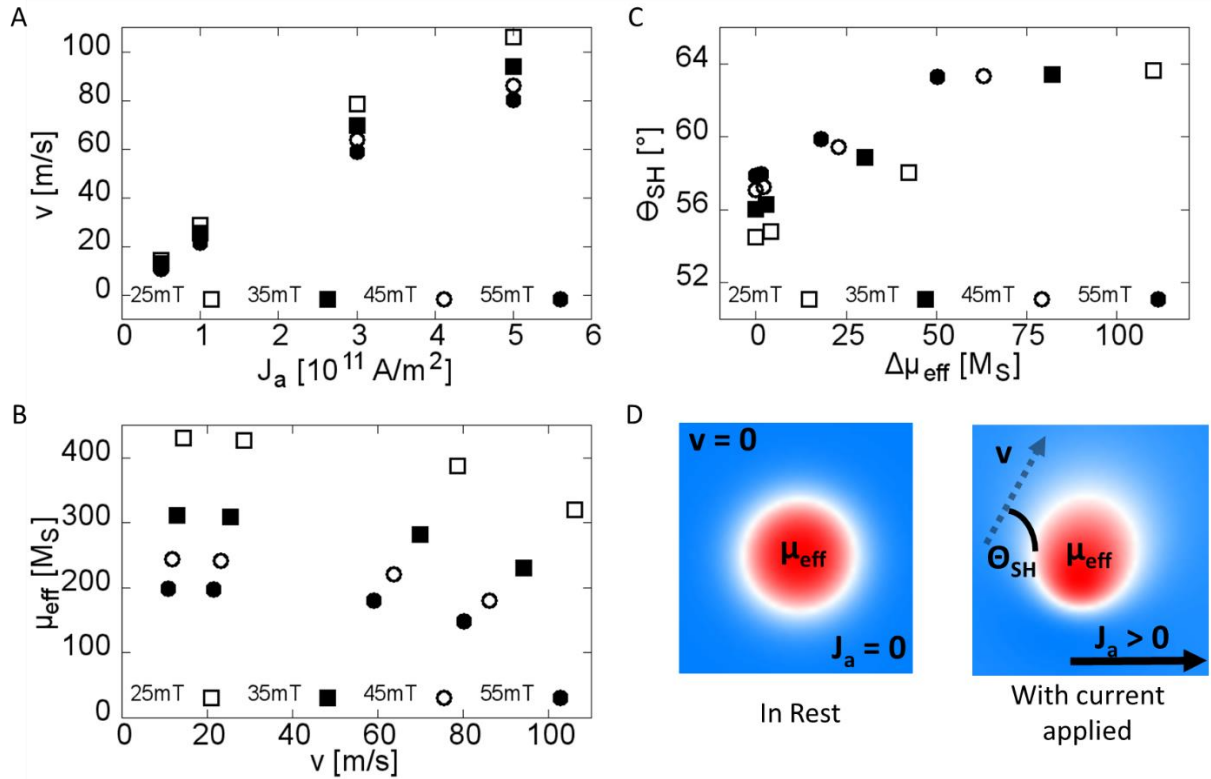

**Figure 4: The FL-SOT as origin of the varying skyrmion Hall angle** simulated with DL- and FL-SOTs ($\xi = 5$) at different out-of-plane fields. A) Dependence of the skyrmion velocity on the current density. B) Dependence of the total magnetic moment $\mu_{eff}$ of the skyrmion (effectively its size) on the velocity. $\mu_{eff}$ is defined as $\mu_{eff} = \sum_{i \,|\, M_{z,i} > 0} \frac{M_{z,i}}{M_S}$, where $M_{z,i}$ is the z-component of the magnetization in cell i and $M_S$ is the saturation magnetization. In this particular case, the skyrmion size decreases slightly with increasing velocity. The change of the skyrmion size entails also a deformation of its shape and can therefore be seen as a measure of this deformation (see D). C) Dependence of the skyrmion Hall angle on the skyrmion's inner moment showing an increase with increasing deformation (plotted as difference $\Delta\mu_{eff}$ of the moment (size) at rest minus the corresponding value with applied current). This indicates that the Hall angle of the skyrmion trajectory is indeed affected by the skyrmion distortion and therefore by the velocity. D) Typical distortion of the skyrmion shape due to the applied current, showing that while the overall area is not strongly affected the shape and the inner region does change.

## SUPPLEMENTARY INFORMATION
## Section 1: DMI Extraction

We determine the DMI strength in the multilayer stack using our established approach of investigating the domain width as a function of the applied out-of-plane (OOP) magnetic field[17]. While approaching the high field regions of the hysteresis loop, the domains of opposite polarity with respect to the applied field approach a terminal width before disappearing. The measured data can be found in Figure S1, which show black/white stripe contrast (corresponding to up and down domains) as well as the periodicity. The terminal width as well as the stripe periodicity were fitted by the function w(x) = a·tanh(ω·x+φ)+d, which can be used to describe a hysteresis loop and is also shown in Figure S1.[31] The finite widths were afterwards extracted from $w_{term} = |d - a|$ and result in values of $w_{term} = 127 \pm 5$ nm and a periodicity of $w_{avge} = 344 \pm 10$ nm. The results of the fitting can be seen in Figure S1 and yield a DMI constant of |D| = (1.35 ± 0.05) mJ m⁻² calculated according to [17]:

$$\frac{\sigma_{DW}}{\mu_0 M_s^2 t} = \frac{w_{aver}^2}{t^2} \sum_{odd\, n=1}^{\infty} \frac{1}{(\pi n)^3}\left[1 - (1 - 2\pi n t / w_{aver})\exp(-2\pi n t / w_{aver})\right]$$

$$\frac{2\pi \sigma_{DW}}{\mu_0 M_s^2 t} = \ln\left[1 + (w_{term}/t)^2\right] + (w_{term}/t)^2 \ln\left[1 + (w_{term}/t)^{-2}\right]$$

$$\sigma_{DW} = 4\sqrt{AK_{eff}} - \pi|D|$$

Here, $\sigma_{DW}$ is the DW energy, M$_S$ the saturation magnetization, t the thickness of the film, A the exchange constant, and K$_{eff}$ the effective uniaxial anisotropy. The high DMI value extracted in this way favors small chiral skyrmions. We use the following nomenclature and definition for the value of the DMI constant D that we provide (note that this is ambiguous in literature):

$$E_{DMI} = -\frac{\mu_0}{2}\int \vec{M}(\vec{r})\vec{H}_{DMI}(\vec{r})dV$$

$$\vec{H}_{DMI} = -\frac{2}{\mu_0 M_S^2}(\vec{D}\vec{\nabla})\times \vec{M}(\vec{r}) \quad \text{with} \quad \vec{D} = \begin{pmatrix} 0 & -D & 0 \\ D & 0 & 0 \\ 0 & 0 & 0 \end{pmatrix}$$

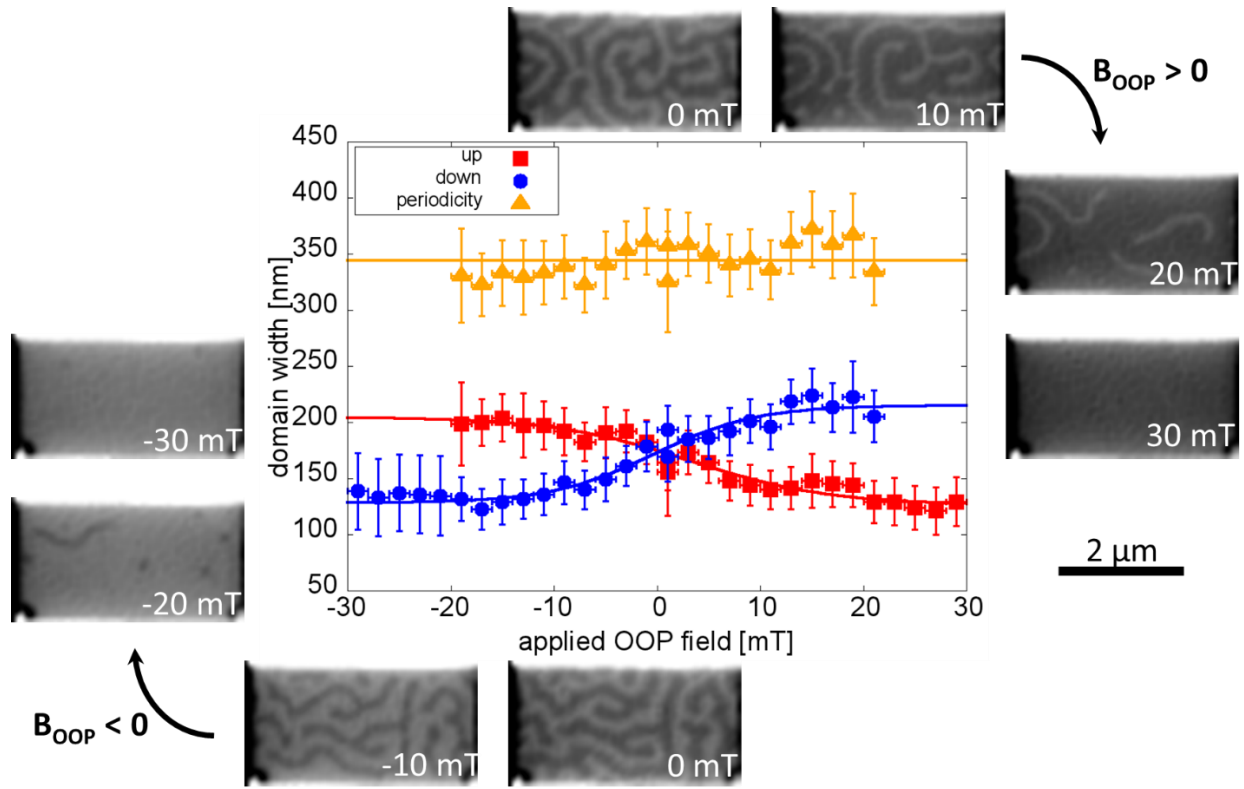

**Figure S1. Domain width as a function of applied OOP field.** The fits were carried out using A·tanh(w·x+c)+d [31] and yielded a terminal domain width of approximately 127 ± 5 nm. The averaged periodicity is 344 ± 10 nm. The DMI can then be extracted using the method established in [17].

**Section 2: Expected Pulse Shapes**

One intrinsic property of the pump-probe technique is the stroboscopic type of imaging, which one has to take into account when predicting the theoretical shape of the skyrmion velocity as a function of current density. For this example, we used a Gaussian excitation pulse and the experimental time step of 2 ns with a time resolution of 70 ps to predict the expected response in the experiment. The result is slightly dependent on the start of the sampling with respect to the start of the pulse. The two plots in Figure S2 show the expected response as a function of time and are in good agreement with the experimental observations. This implies that the experimental system can be described well by micromagnetic simulations.

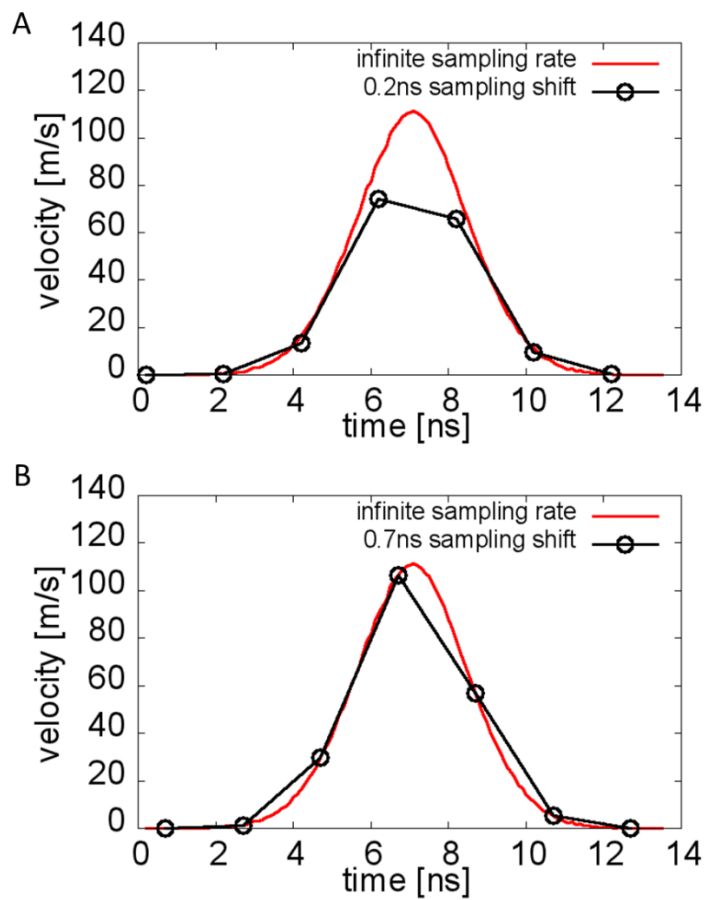

**Figure S2. Influence of sampling rate on expected skyrmion velocities** as calculated by micromagnetic simulations with approximated Gaussian current pulse shape. The peak current density of $4 \times 10^{11}$ Am$^{-2}$ corresponds to the experimentally used one in Figure 2D. The two plots correspond to different relative sampling delays (A = 0.2 ns; B = 0.7 ns). The peak shapes show excellent agreement with the experimental data.

### Section 3: Influence of Materials Parameters on the Skyrmion Hall Angle

One possible explanation for the varying skyrmion Hall angles as a function of current density could be that heating effects or imprecisely measured material parameters may affect the trajectories. To exclude such uncertainties and variations in the materials parameters as the origin of the observed skyrmion Hall angles, we show here the influence of up to 20% deviation of the parameters from the experimentally measured ones. Figure S3 shows that there is a slight influence on the angles, however varying conventional materials parameters, e.g. due to heating effects, cannot explain the strong observed dependence of the skyrmion Hall angle on the velocities.

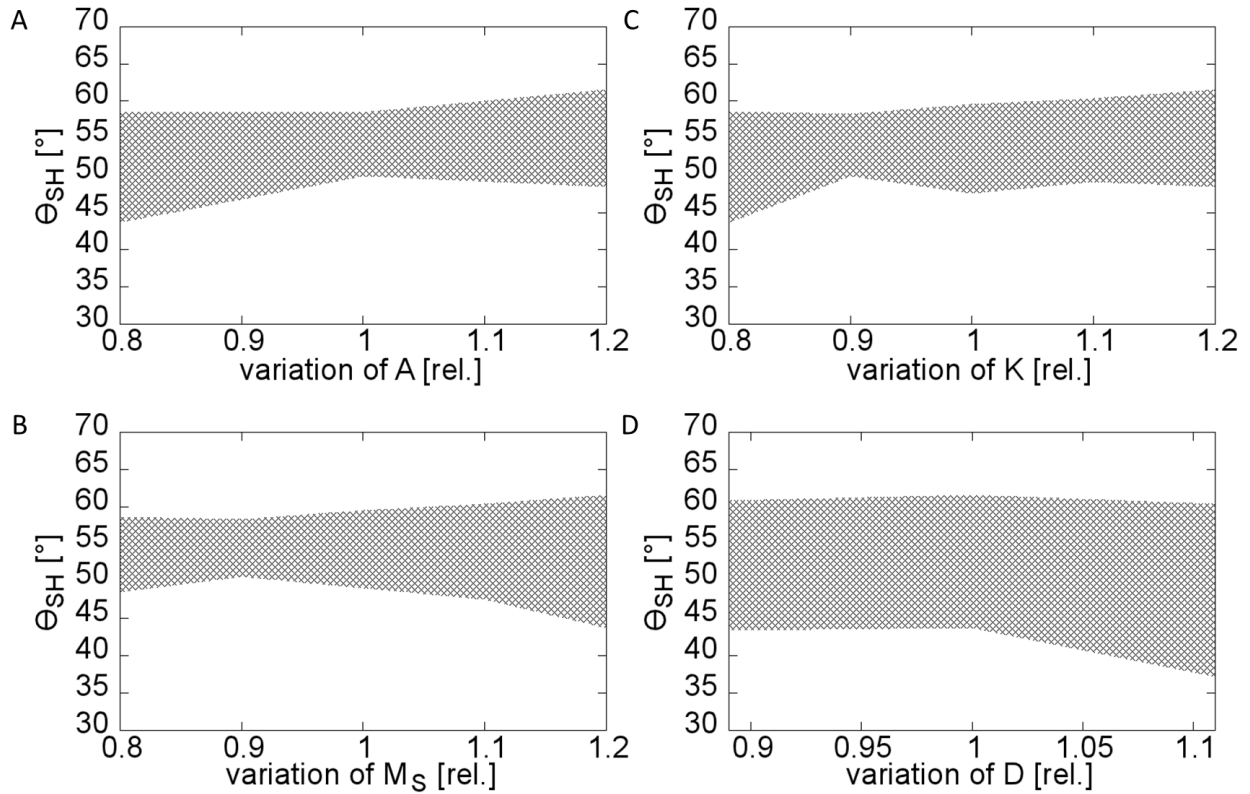

**Figure S3. Estimation of the variation of the skyrmion Hall angle due to a change in material parameters.** The estimation was calculated micromagnetically by varying anisotropy, exchange, saturation magnetization, and DMI up to 20% from the experimentally observed values. Shown is the 3σ-interval of the estimated change of the skyrmion Hall angle sorted by: A) varying exchange constant A; B) varying anisotropy K; C) varying saturation magnetization $M_s$; D) varying DMI constant D. We see that the parameter changes do not lead to large changes in the skyrmion Hall angle. The 1σ-interval can be estimated by ±5°. The only parameter that has a slightly higher influence is the DMI for a change of >0.1 mJm$^{-2}$. However, there is no evidence that the DMI exhibits a strong temperature dependence for the temperature range above room temperature that we probe by current heating and as heating is proportional to $I^2$, the observed linear dependence of the angle on the velocity cannot be explained based on changes of these materials parameters.

## Section 4: Influence of the FL-SOT on the Skyrmion Hall Effect

To demonstrate that the FL-SOT has a strong influence on the skyrmion behavior, we simulated the skyrmion Hall angle as function of the current density for different values of the FL-constant ξ as well as the influence of ξ directly on the Hall angle. As visible in Figure S4 and S5, ξ indeed influences the behavior of the skyrmions systematically. While a pure DL-SOT (ξ = 0) shows no significant dependence of the angle as function of the current density, finite values of ξ can lead to a strong current-dependence of the skyrmion Hall angle. Since the sign of the dependence (increase or decrease) depends on the sign of ξ, this allows to determine the sign of the FL-SOT in these systems. While the absolute angles obtained here do not quantitatively agree, this shows that the deformation of the skyrmions together with the FL-SOT can definitely lead to a dependence of the skyrmion Hall angle on the current density as observed in the experiment. To obtain quantitative agreement, finite temperature simulations are needed, which better reproduce the excitations of the skyrmion. These are, however, beyond the scope of this work and not widely accessible.

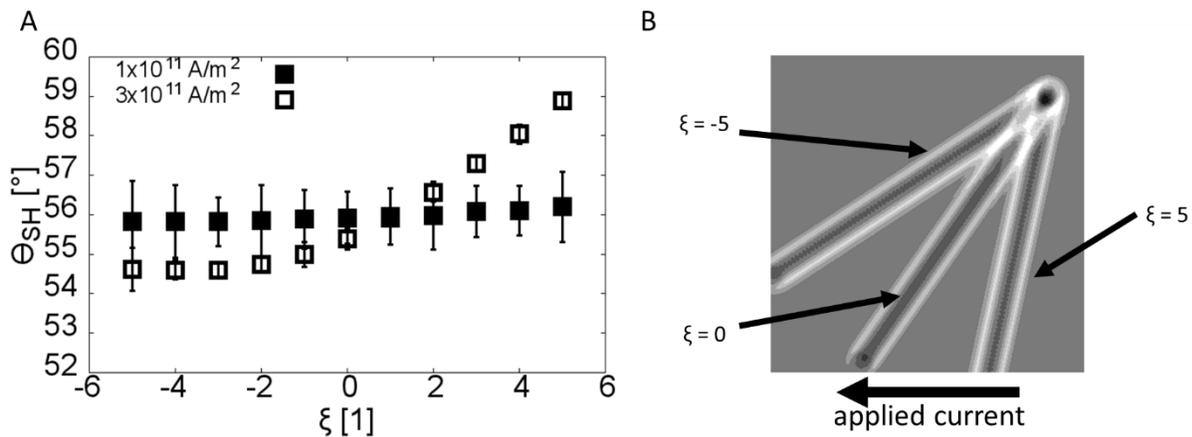

**Figure S4. Calculated skyrmion Hall angles for different FL-SOTs.** A) Dependence of the skyrmion Hall angle on the FL-SOT at two different current densities. The value of ξ shows an increasing influence for higher current densities. B) shows a schematic of the skyrmion trajectories three values of ξ. Negative values tend to reduce the angle while positive ones increase it.

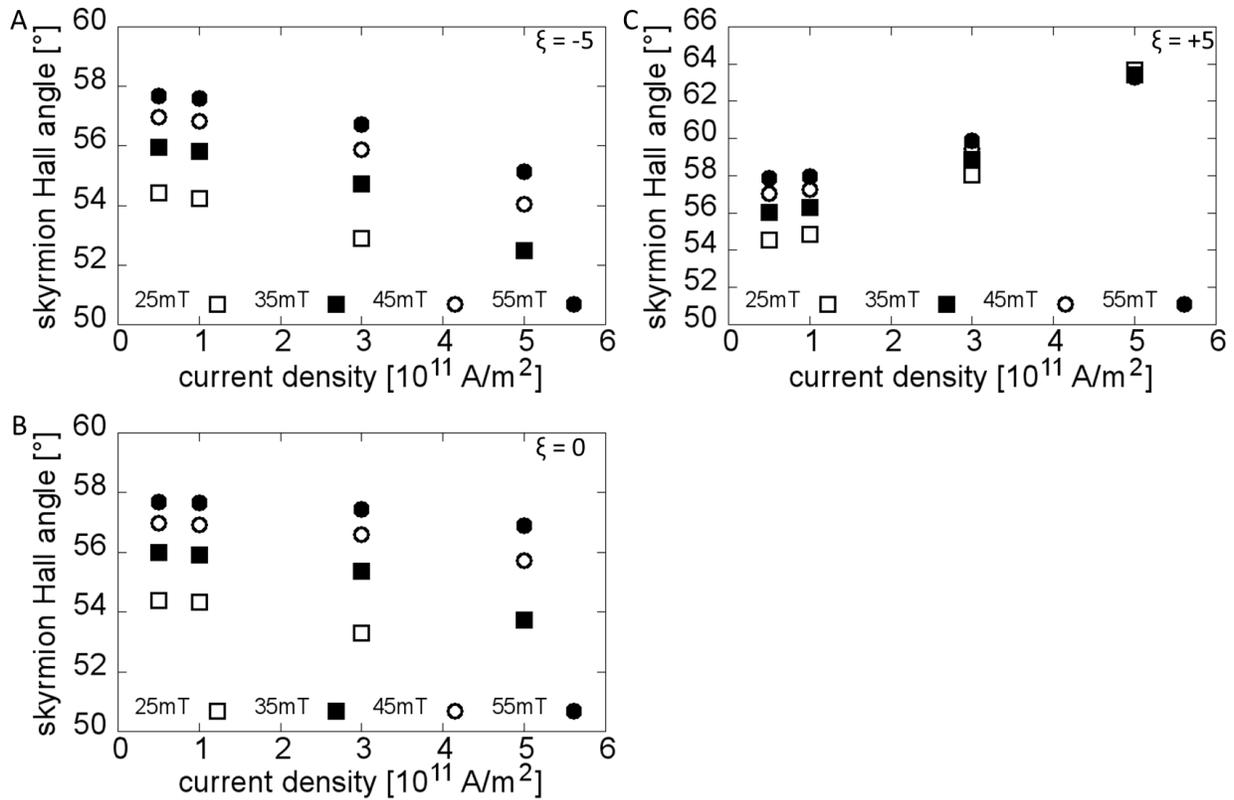

**Figure S5. Calculated skyrmion Hall angle dependence on the applied current density for different FL-SOTs.** Shown is the dependence of the skyrmion Hall angle on the current density for $\xi = -5$ (A), $\xi = 0$ (B) and $\xi = 5$ (C). For only a DL-SOT ($\xi = 0$) no significant dependence of the skyrmion Hall angle on the current density was found.

**Section 6: Experimental pump-probe movie**

An example of the experimentally obserbed skyrmion dynamics (current density $3.5 \times 10^{11}$ A/m$^2$) can be found attached to this supplementary information. In this time-resolved movie, the real-time behavior of several skyrmion in our system can be seen. Several skyrmions and their trajectories are indicated as overlay. The current pulses are marked in the upper right corner.